\documentclass[letterpaper,aps,pra,showpacs,twocolumn,groupedaddress]{revtex4-1}
\usepackage{graphics,color,epsfig}
\usepackage{amsmath,amssymb}

\newcommand{\ka}{\hbox{\ae}}
\begin{document}
%\setboolean{shortarticle}{true} % true = letter, false = research article
\title{Tree-wave mixing of ordinary and backward electromagnetic waves: extraordinary transients}

\author{Vitaly V. Slabko$^{1}$, A. K. Popov$^{2,*}$, Viktor A. Tkachenko$^{1}$, S. A. Myslivets$^{3,1}$}
\affiliation{
$^1$Siberian Federal University, 79 Svobodny  Av., Krasnoyarsk, 660041, Russian Federation\\
$^2$Birck Nanotechnology Center, Purdue University,
West Lafayette, IN 47907,~USA\\
$^3$L. V. Kirensky Institute of Physics, Siberian Division of the Russian Academy of Sciences, 660036 Krasnoyarsk, Russia \\
$^*$Corresponding author: popov@purdue.edu}

%\author[1]{Vitaly V. Slabko}
%\author[2,*]{Alexander K. Popov}
%\author[1]{Viktor A. Tkachenko}
%\author[3,1]{Sergey A. Myslivets}
%
%\affil[1]{Siberian Federal University, 79 Svobodny  Av., Krasnoyarsk, 660041, Russian Federation}
%\affil[2]{Birck Nanotechnology Center, Purdue University; 1205 W State Str., West Lafayette, IN 47907, USA}
%\affil[3]{L. V. Kirensky Institute of Physics, Siberian Branch of the Russian Academy of Sciences, Krasnoyarsk, 660036, Russian Federation}
%
%\affil[*]{Corresponding author: popov@purdue.edu}
%
%\dates{Compiled \today}

%\ociscodes{(190.4223)   Nonlinear wave mixing; (190.4410)   Nonlinear optics, parametric processes; (190.5530)   Pulse propagation and temporal solitons; (160.3918)   Metamaterials.}

%\doi{\url{http://dx.doi.org/10.1364/ol.XX.XXXXXX}}

\begin{abstract}
Three-wave mixing of ordinary and backward electromagnetic waves in pulsed regime is investigated in the metamaterials, which enable co-existence and phase matching of such waves. It is shown that opposite direction of phase velocity and energy flux in backward waves gives rise to extraordinary transient processes in greatly enhanced optical parametric amplification and in frequency up or down shifting nonlinear reflectivity. The discovered transients resemble slowed response of an oscillator on pulsed excitation in the vicinity of its resonance.
\end{abstract}

%\setboolean{displaycopyright}{true}

\maketitle
%\thispagestyle{fancy}
%\ifthenelse{\boolean{shortarticle}}{\abscontent}{}

Advances in nanotechnology have made possible  engineering of the metamaterials (MMs) which  support backward optical electromagnetic waves (BEMWs). Counter-intuitively, energy flux and phase velocity are \emph{contra-directed} in BEMWs. Such waves may exist only in a certain frequency range dependent on particular MM, whereas only ordinary EMWs propagate outside the indicated frequency band. The emergence of BEMWs gave rise to revolutionary breakthroughs in linear photonics \cite{ShC}. Exciting possibilities were predicted in nonlinear photonics, such as huge enhancement in wave mixing  through implementation of BEMW as one of the coupled waves \cite{ShC, Popov:Spr.193-213.2015}. Predicted extraordinary  enhancement in  coherent energy  transfer from the pump wave to \emph{contra-propagating} electromagnetic  waves at different frequencies opens broad prospects for practical use of this phenomenon in photonics. Among them are of optical parametric amplification (OPA) and compensating losses, cavity-less parametric oscillators, modulators  and amplifiers, as well as all optically controlled, frequency up and down shifting nonlinear optical reflectors and sensors  in optical and microwave ranges of electromagnetic radiation \cite{ShC, Popov:Spr.193-213.2015}.
Only few realizations of coherent coupling of contra-propagating optical waves have been reported so far \cite{Mw,Such,Canalias:NatPhot.1.459}. Current mainstream in engineering of MMs which support BEMWs is to craft the MMs composed of plasmonic mesoaatoms, which introduce negative $\mu$ and $\epsilon$ (negative-index MMs, NIMs) \cite{ShC}. A more broad class of MMs, which can support coexistence of \emph{ phase-matched} ordinary and BEMWs, was recently proposed based on negative spatial dispersion and nanowaveguide operation regimes \cite{APA2012,AST2013,Popov:SPIE.8725,SSP2014,CNT,Duncan}.
\begin{figure}[htbp]
\centering
\includegraphics[width=0.48\columnwidth]{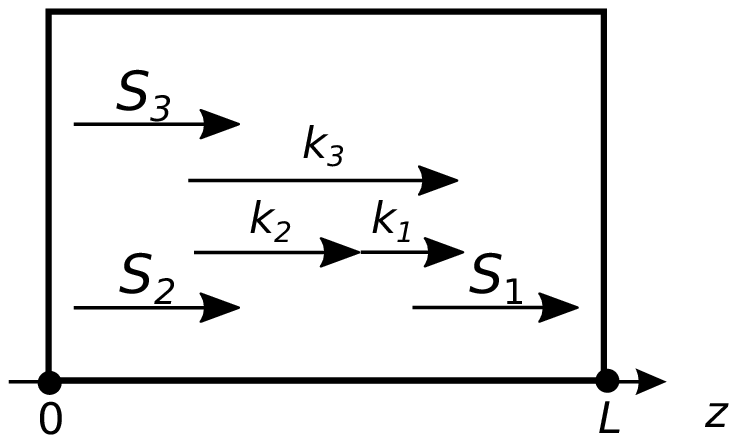}
\includegraphics[width=0.48\columnwidth]{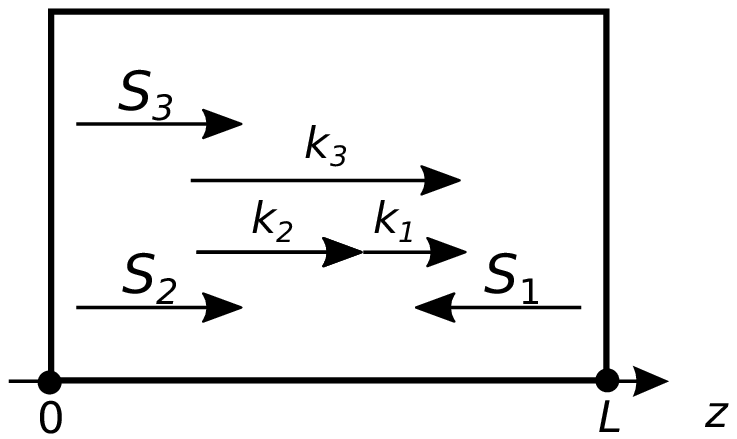}
\\
(a)\hspace{0.5\columnwidth} (b)
\caption{Two alternative coupling options: (a) -- co-propagating waves; and (b)-- contra-propagating signal and backward-wave idler waves. Here, $ \vec {k} _j $ are wave vectors and  $ \vec{S}_j $  are Poynting vectors.}
\label{f1}
\end{figure}
A remarkable difference in the nonlinear optical (NLO) three-wave mixing (TWM)  $\omega_1= \omega_3-\omega_2$ in the alternative cases of co- and contra-propagating waves at frequencies $\omega_1$ and $\omega_2$ is as follows. For co-propagating EMWs coupled in an ordinary loss-free material [Fig.\ref{f1}(a)], a signal at $\omega_2$ grows as $a_2(L)\sim \exp (gL)$. However, this dependence dramatically changes to   $a_2 (L)\sim 1/\cos (gL)$ in the case of Fig.~\ref{f1}(b) where  ordinary wave at $\omega_2$ propagates along the pump wave at $\omega_3$ whereas the phase matched idler at $\omega_1$ is a BEMW which propagates in the opposite direction  \cite{Bob,Har,APB,OL,EPJD,SPIE}.
\begin{figure}[htbp]
\centering
\includegraphics[width=0.6\columnwidth]{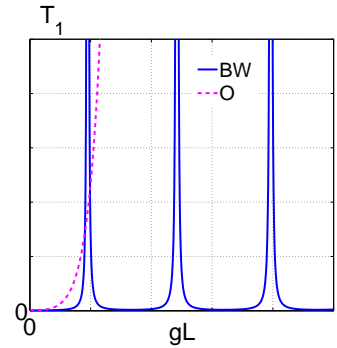}
\caption{Difference in the dependence of the output parametrically amplified signal in the transparent slab on the slab length $L$ and on intensity of the pump field (factor $g$) for ordinary (O) and backward-wave (BW) settings.}
\label{f2}
\end{figure}
 Here, $a_2$ is amplitude of the transmitted signal wave, $L$ is thickness of the NLO slab, $g$ is a factor proportional to NLO susceptibility and to amplitude $a_3$ of the pump wave at $\omega_3$. It is seen that, in the latter case,  \emph{transmittance} $T_2$  at $\omega_2$, (\emph{OPA}) experiences \emph{extraordinary enhancement} as $gL\to (2q+1)\pi/2$ (q=0, 1, 2...). The output idler at $\omega_1$ contra-propagating  in the \emph{reflection} direction (NLO \emph{frequency up or down shifted reflectivity}) experiences similar huge, resonant, enhancement.  Such "geometrical" resonance dependence of the TWM output on $L$ (or on $g$) occurs due to nonlocal, of distributed feedback type,  NLO coupling intrinsic to the BW setting.  The outlined resonance dependence depicted in Fig.\ref{f2} resembles resonance property of a cavity composed of two mirrors as the cavity length L or wave vector $k_q$  approach resonance values $k_qL=(2q+1)\pi/2$ (q=0, 1, 2...). The latter  occurs as two coherent contra propagating waves meet each other  with equal phases to interfere constructively at a given point inside the cavity. Oscillation resonances are known to cause transient processes under pulsed excitation, which is the behavior  intrinsic to any oscillator.

 This paper is to demonstrate such unparalleled transient processes in OPA and in the NLO  reflectivity   in the vicinity of the resonance intensities of the pump (control) field  at $\omega_3$. Such transients are \emph{extraordinary} because they are not inherent to TWM  in ordinary  materials. The outlined NLO reflectivity has no analog in ordinary materials at all.

Consider  interaction of three electromagnetic waves $ E_j (z, t) = A_j(z, t) \exp\{i (\omega _j t-k_j z)\} $ (j=1, 2, 3) in a loss-free medium of length $L$ with quadratic nonlinearity $\chi ^{(2)}$. Wave vectors of all waves are co-directed along  the axis $ z $ . The relations  $ \omega _3 = \omega _1 + \omega _2 $,  $ k_3 = k_1 + k_2 $ and $ \left ({A_3 >> \, A_1, \, A_2} \right) $ are supposed met. After introducing
 amplitudes $e_{j}=\sqrt{|\epsilon_j|/k_j}A_j$,  $a_j=e_i/e_{10}$, coupling parameters $\ka=\sqrt{k_1k_2/|\epsilon_1\epsilon_2|} 4\pi\chi^{(2)}_{\rm eff}$ and $g=\ka A_{10}$, where $ \varepsilon _1$ and $\varepsilon _2 $ are dielectric permittivity of the medium at the corresponding frequencies, $A_{j0}=A_j(z=0)$,   equations for   electric components of waves in the approximation of slowly varying amplitudes can be written as \cite{CNT,EPJD,SPIE}:
\begin{equation}
\label{eq1}
\left\{ {\begin{array}{l}
{\partial a_1}/{\partial z}+({1}/{v_1}){\partial a_1}/{\partial t}=
     -s_1iga_3a_2^*, \\
 {\partial a_2}/{\partial z}+ ({1}/{v_2}){\partial a_2}/{\partial t}=
  -iga_3a_1^*, \\
 {\partial a_3}/{\partial z}+({1}/{v_3}){\partial a_3}/{\partial \tau}=
     -iga_1a_2.
 \end{array}} \right.
\end{equation}
Here, $v_j$ are group velocities at the corresponding frequencies.  For  ordinary wave at $\omega_1$ [Fig.~1(a)], $v_1>0$, $s_1=1$. For BW at $\omega_1$ [Fig.~1(b)], $v_1<0$, $s_1=-1$. Quantities  $|a_j|^2$ are proportional to the time dependent photon fluxes, $g^{-1}$ is a characteristic slab length required for significant NLO energy transfer from the pump field $A_3$ to the signal at $\omega_2$ and to the idler at $\omega_1$. In the case of Fig.~1(a), the boundary conditions are defined as  $ a_1 (z = 0) = 0 $, $ a_2 (z = 0) = u $. In the case of Fig.~1(b), they  must be written as  $ a_1 (z = L) = 0 $ and $ a_2 (z = 0) = u $.  Here, $u\ll a_{30}$.  It is readily seen, e.g. for the ultimate  case of steady state regime and constant $a_3$,   that the indicated changes give rise to fundamental transformation of solution to the coupled  Eqs.~(\ref{eq1}), which predict  appearance of the outlined geometrical resonances instead of exponential dependence (Fig.~2).

 We introduce transmission (OPA) factor $ T_2=|a_2 (z = L, t) /a_2 (z = 0) |^2 =|a_{2L}(t) / a_ {20}|^2$, and the NLO reflectivity factor $ R_1=|a_1 (z = 0, t) / a_ {20}|^2=|a_{10}(t) / a_ {20}|^2$.
In further investigations, the signal wave $A_2$ and the difference-frequency idler $A_1$ are supposed to be so weak and energy transfer small that the depletion of the fundamental wave $A_3$ can be neglected. This does not preclude huge amplification of a week signal.

Basically, two different regimes are possible. In the first one, $E_2$ is a semi-infinite rectangular pulse and $A_3$ is a continuous wave (cw). This regime will be referred to as the \emph{amplification mode}. The opposite case of a cw  $A_2$ and a semi-infinite rectangular pulse of $E_3$  will be referred to as the \emph{generation mode}.

First, consider the case of $v_3=v_2=-v_1$. Shape of semi-infinite pulse with sharp front edge travelling with group velocity $v_3$ along the axis $z$ will be depicted by function $F(t) =\left\{1- \tanh\left[(z / v_3 -t) / t_f \right] \right\}/2$, where parameter $t_f$ determines its edge steepness. In the following numerical simulation, it is taken equal to $ t_f = 0.05\Delta t$, where $\Delta t= L /v_3$ is a travel time of the pump pulse front edge through the slab.
In the amplification mode, $a_2(t, z=0)= a_{20}F(t)$, $a_3(t)= a_{30}$= const. In the generation mode,  mode, $a_3(t, z=0)= a_{30}F(t)$, $a_2(t, z=0)= a_{20}$= const. Here, $ a_ {j0} $  is a maximum pulse amplitude value at the slab entrance. Solution to the Eqs.~(\ref{eq1}) was obtained through numerical simulations.
\begin{figure}[htbp]
\centering
\includegraphics[width=0.49\columnwidth]{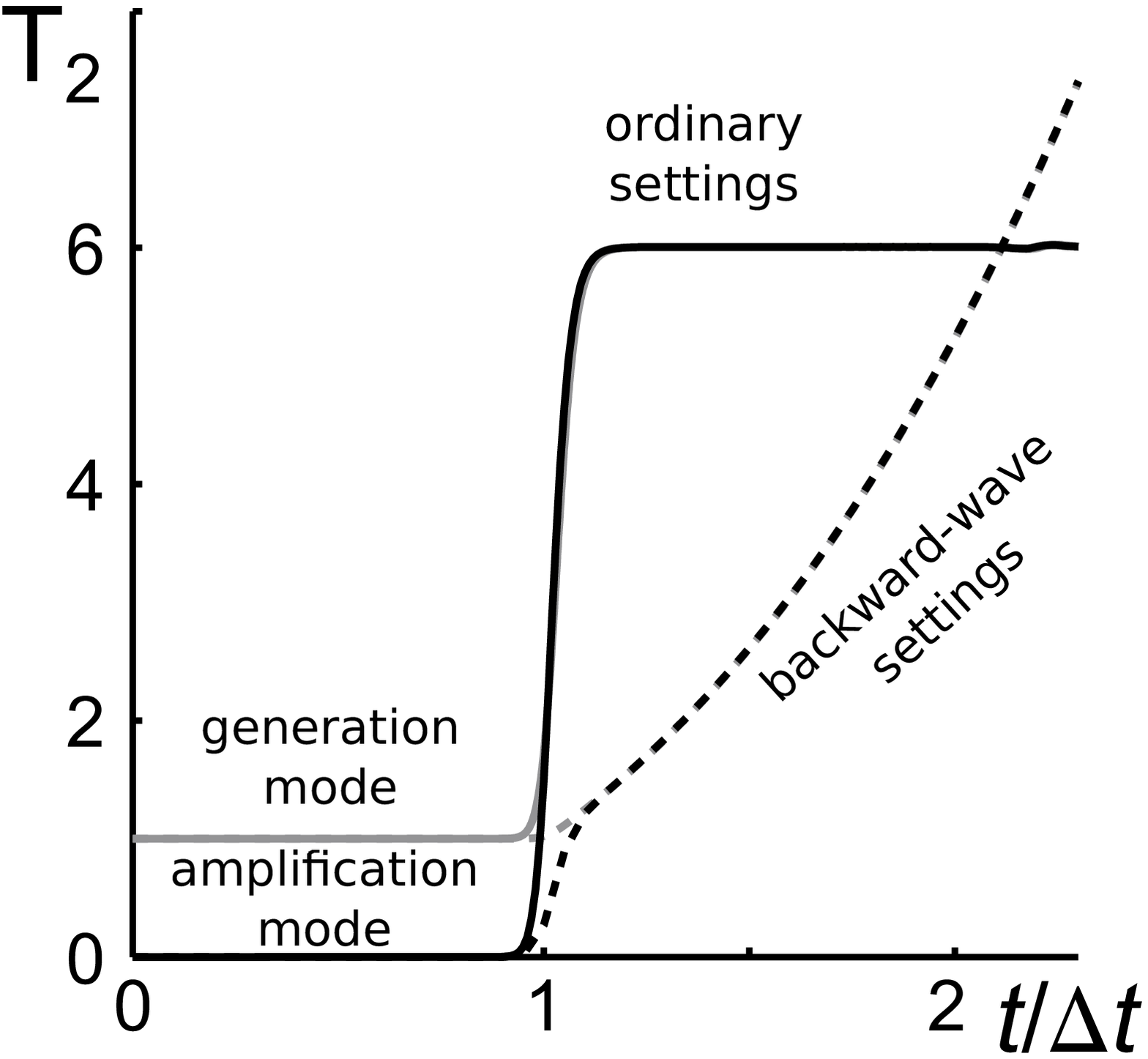}
\includegraphics[width=0.49\columnwidth]{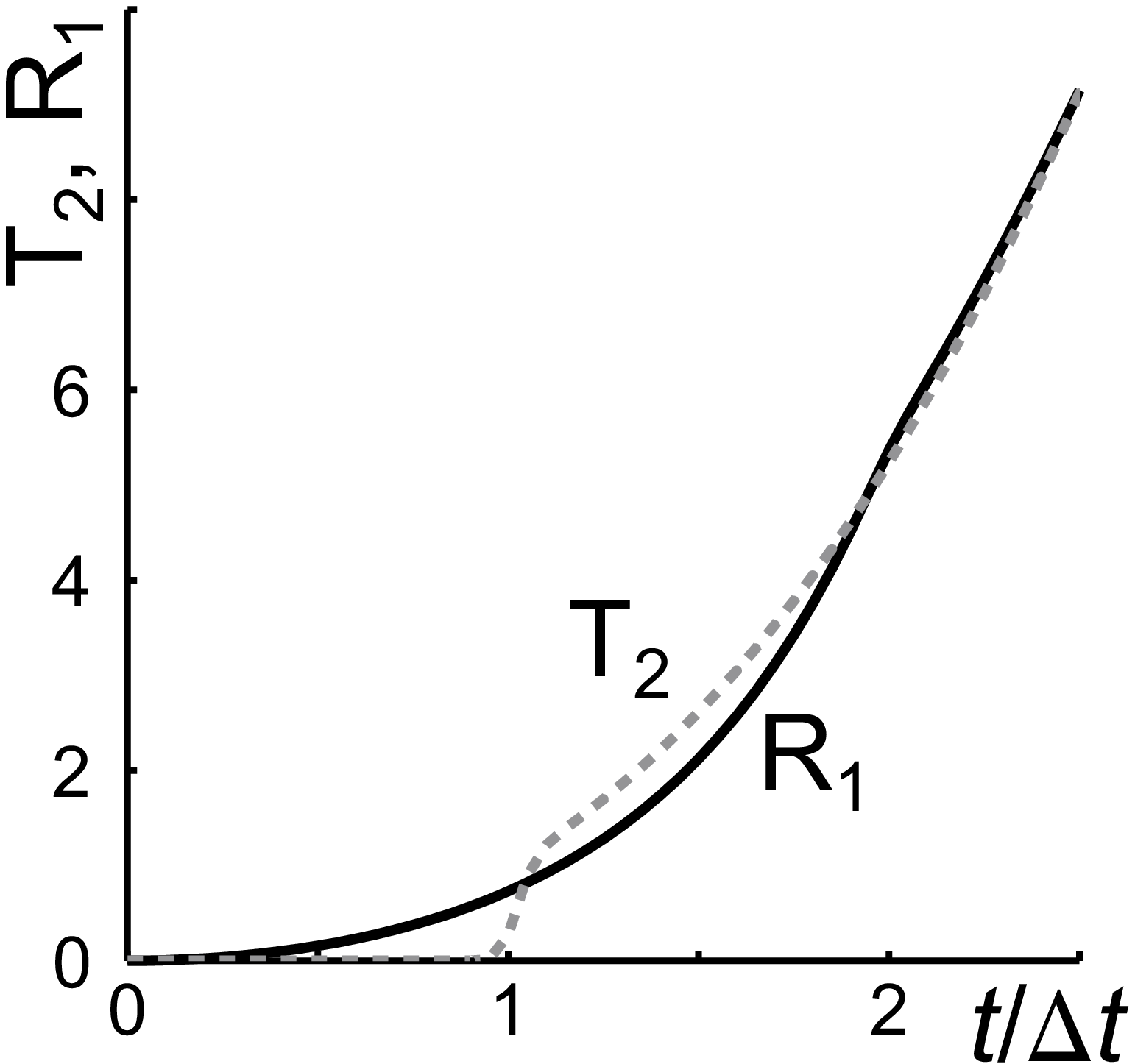}\\
(a)\hspace{0.5\columnwidth}(b)
\caption{(a) Difference in the transient processes under  ordinary (the solid lines) and BW (the dashed lines) settings.  $T_2(t)$  is transmission (OPA) of the co-directed seeding signal at the forefront of the output signal pulse.
(b) Difference in the transient processes in  $T_2(t)$ and in  nonlinear-optical reflectivity $R_1(t)$ (contra-propagating generated idler) vs time in the amplification mode.  (a) and (b): $v_3=v_2=-v_1$, $gL=0.984\pi/2$.}
\label{f3}
\end{figure}

Figure~\ref{f3}(a), the solid line, depicts output signal at the slab exit $z=L$ in the case of co-propagating ordinary waves and the dashed line -- in the case of the BW setting. In the amplification mode, any changes in the output signal occur only after the travel period, both in the ordinary and BW cases.    In the generation mode, the output experience amplification when forefront of the pump pulse reaches the exit and it almost follows shape of the pump pulse in the ordinary coupling. However, in the BW setting, pulse shape changes dramatically in the vicinity of the resonance intensity of the pump wave. The growth occurs slower; the output signal is greater and its maximum is reached with significant delay.

Figure~\ref{f3}(b) shows the transmitted signal at $z=L$ in the amplification mode and the generated in the reflection direction idler at $z=0$. It takes the travel time $\Delta t$ for the  signal pulse to appear at $z=L$, whereas the idler is generated immediately after the pulse enters the slab. Hence, unlike the transmittance, the transients in the reflectivity are the same in the amplification and in the generation modes.

Figure~\ref{f3} is to depict time dependence of the signal and the idler  within small time interval about $2\Delta t$ after the pulse enters the slab.  It shows that differences in OPA in the  generation and in the  amplification   modes disappear after period of time about $\Delta t$, whereas  the reflectivity and OPA  develop in a similar way only after period of time about $2\Delta t$.
\begin{figure}[htbp]
\centering
\includegraphics[width=0.8\columnwidth]{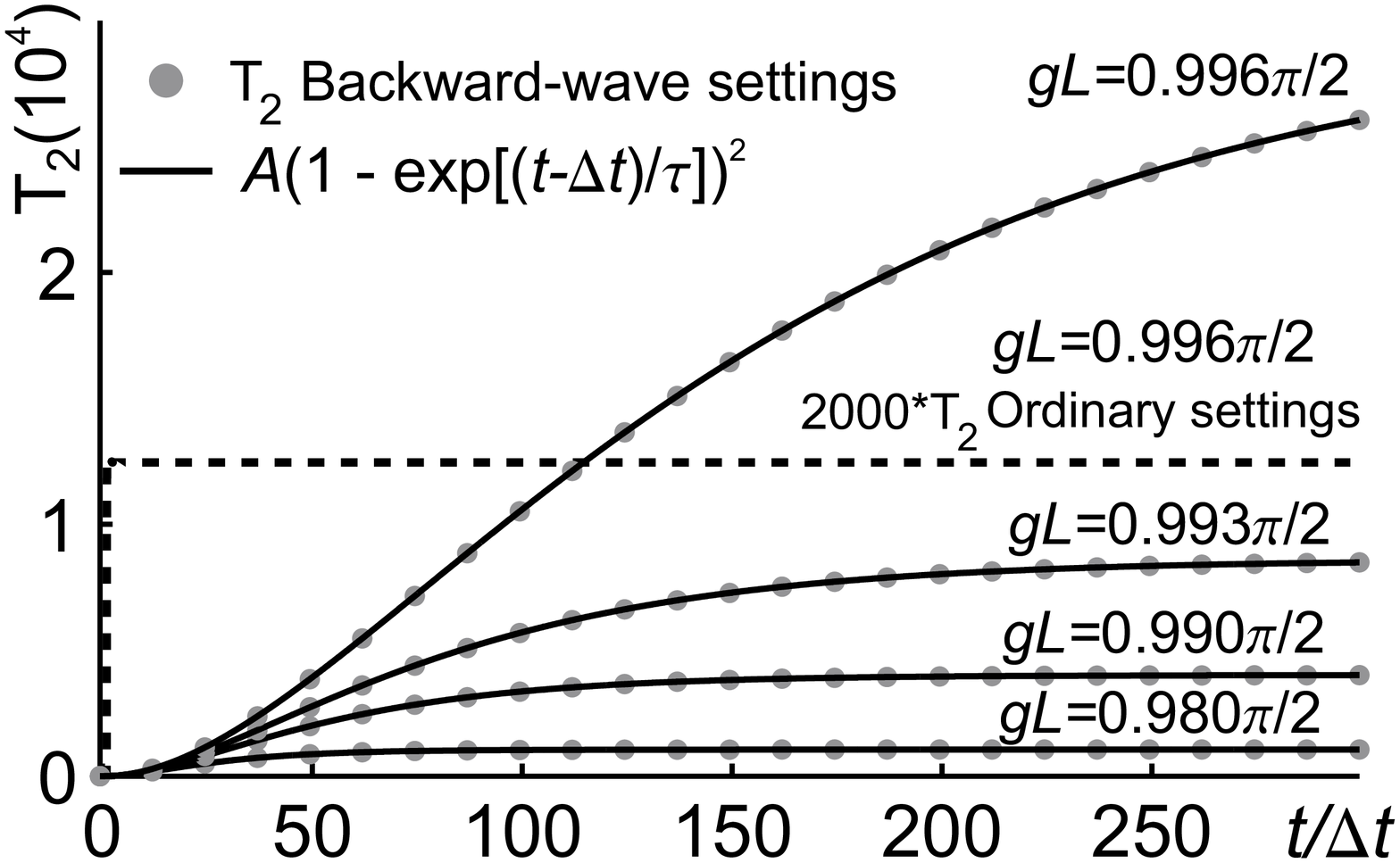}\\
\caption{ Dependence of the transient OPA on the intensity of the pump field.
 The dashed line -- TWM of ordinary waves. Points correspond to ordinary signal and contra-propagating idler.  The solid lines depict the approximation of the transient OPA by the function $T_2(t)=A(1-\exp{[(t-\Delta t)/\tau]})^2$.   $v_3=v_2=-v_1$.}
\label{f4}
\end{figure}

Figure~\ref{f4} demonstrates that the rise time and the maximum of OPA grow with approaching the resonance strength by the pump field. It also demonstrates fundamental difference between the rise periods and  maxima achieved in the ordinary and BW TWM. It is seen that the rise period may reach impressive values on the order of hundred travel periods $\Delta t$. It appears that calculated data can be approximated by the exponential dependence $T_2(t)=A(1-\exp{[(t-\Delta t)/\tau]})^2$  (the solid lines), where the rise time $\tau$ grows approximately as $1/\cos(gL)$ with approaching the intensity resonance. For example, at $gL=0.996\pi/2$, the fitting values are  $A=2.99\cdot 10^4$, $\tau=109.9 \Delta t$. Here, the dependencies blown up in Fig.~\ref{f3} are not resolved.  The transient processes in OPA and in the NLO reflectivity are similar.

\begin{figure}[htbp]
\centering
\includegraphics[width=0.485\columnwidth]{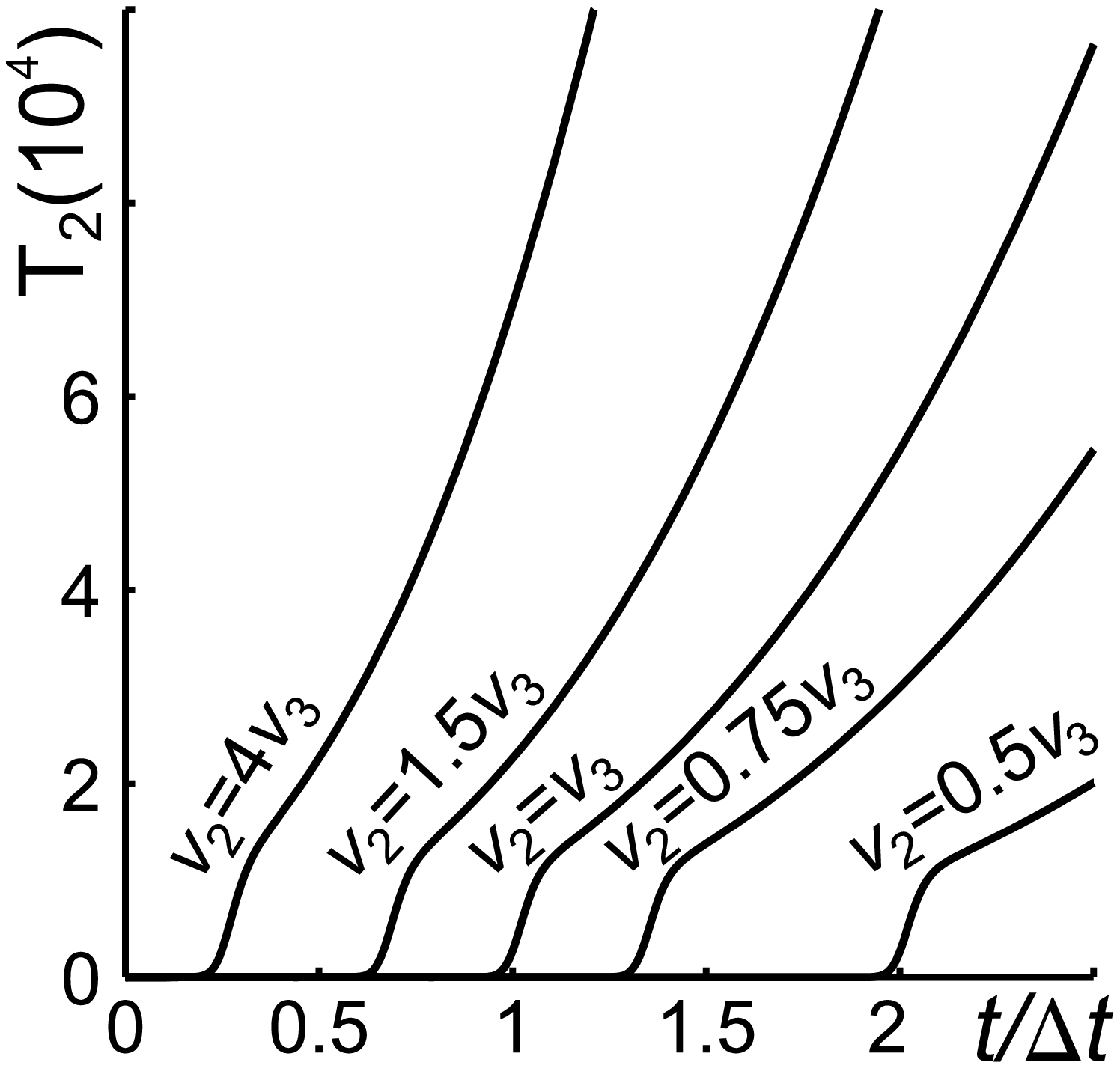}\hspace{1ex}
\includegraphics[width=0.485\columnwidth]{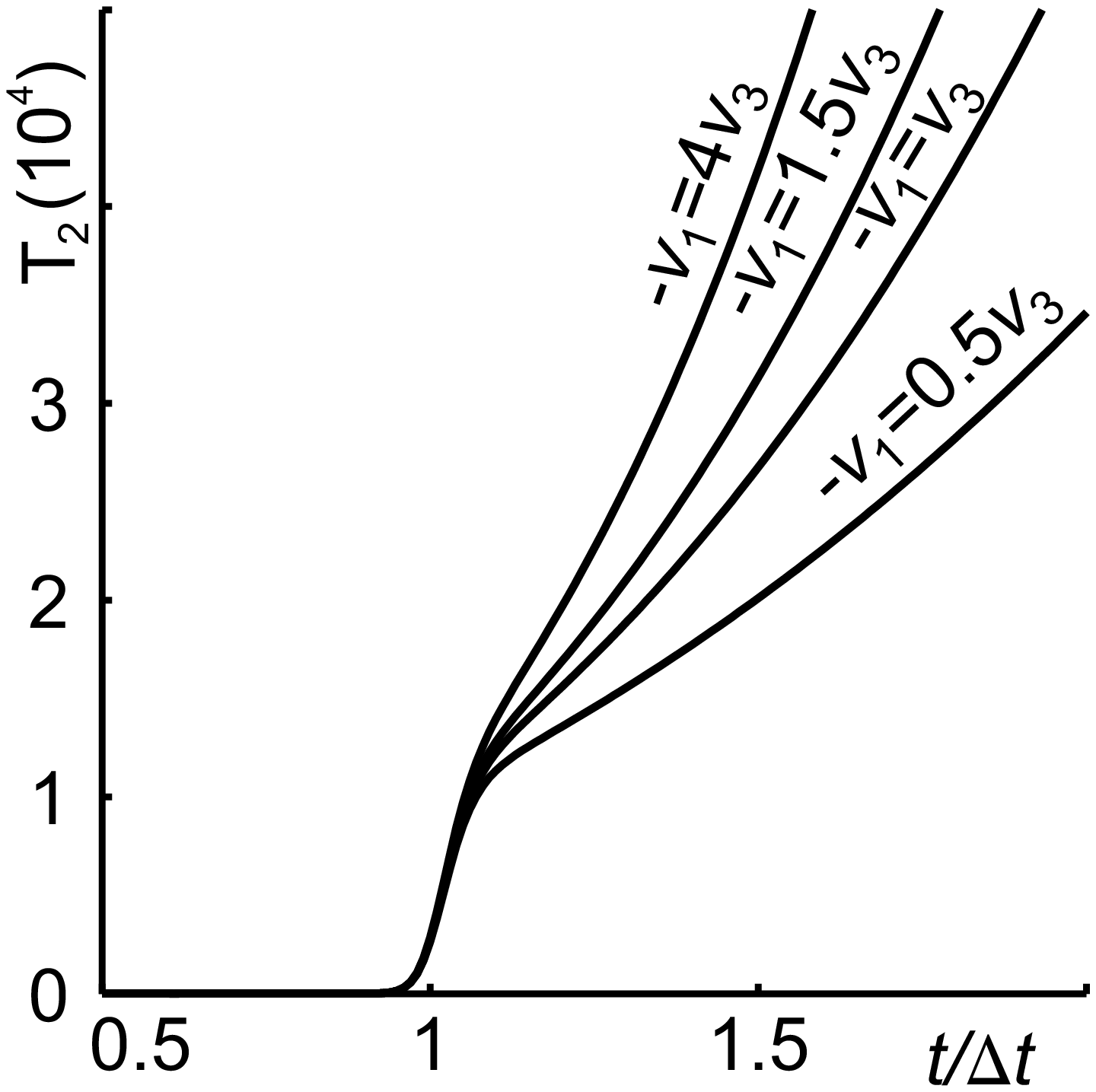}\\
(a)\hspace{0.5\columnwidth}(b)
\caption{Dependence of the transient TWM on  dispersion of  group velocities of the coupled waves. (a): $-v_1=v_3$. (b): $v_2=v_3$. (a) and (b): $gL=0.996 \pi/2$. }
\label{f6}
\end{figure}

\begin{figure}[htbp]
\centering
\includegraphics[width=0.9\columnwidth]{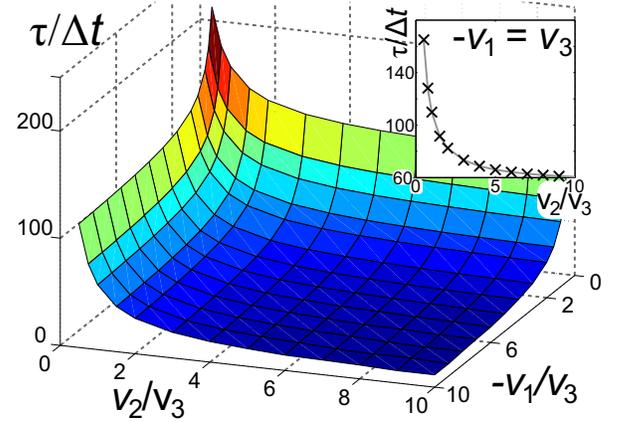}%\\
\caption{Dependence of  duration of the transient period in TWM  on  dispersion of  group velocities of  the coupled waves. $gL=0.996 \pi/2$.}
\label{f7}
\end{figure}

Figures~\ref{f6} (a) and (b)  depict dependence of the transient processes at the forefront of the output signal on group velocity dispersion for small initial time interval. Figure~\ref{f6}(a) shows that the forefront of the output signal becomes steeper if it outruns the pump pulse. Corresponding changes in the rise time are shown in the inset in Fig.~\ref{f7}. Figure~\ref{f6}(b) demonstrates that the pulse forefront also becomes steeper with increase of group velocity for the idler. Figure~\ref{f7} shows that the rise time maximizes at equal modules of group velocities.

In the conclusion, we report extraordinary transient processes in optical parametric amplification and in the frequency-shifted nonlinear optical reflectivity, which originate from coherent nonlinear optical coupling of ordinary and contra-propagting backward  electromagnetic waves.
Energy flux (group velocity) and phase velocity are counter-directed in a backward wave. It is shown that  such an exotic property gives rise to  transient processes in three-wave mixing that have no analogs in the ordinary nonlinear-optical materials.   In order to explicitly demonstrate the difference, the ultimate case is investigated through numerical modeling. It is  semi-infinite pulse with sharp rectangular forefront entering  a loss-free  metamaterial slab, whereas two other coupled waves are  continuous waves travelling with the same phase velocities as the input pulse. The reference is given to the work showing how such a requirement can be realized. Phase matching dictates energy flux in the backward wave (here, for the idler)  to be directed against those in the pump and signal, i.e., in the reflection direction. The latter distinguishes ordinary and backward-wave materials and gives rise to greatly enhanced coherent energy conversion from the pump to the signal and contra-propagating idler and to the  demonstrated fundamental differences in the transient processes.   Two options are investigated where either  incident pump  or co-propagating signal are pulsed. In both cases, output signal and propagating in opposite direction idler appear pulsed.  The results are compared with the similar processes under standard settings, where all coupled waves are ordinary and co-propagating. Major conclusions are as follows. There exist resonance values of the input pump intensity, which is dependent on nonlinear susceptibility and thickness of the metaslab,  that provides giant enhancement in three-wave coupling.  Consequently, great enhancement occurs in optical parametric amplification of the signal and in the oppositely directed idler (in the frequency up or down shifted reflectivity). Such effect does not exist in the ordinary optical parametric amplification in the case of all co-directed energy fluxes, where the indicated reflectivity does not exist at all. In the vicinity of the resonance intensity, extraordinary  transient processes develop which cause a change in the output pulse shape. The latter is not the case under ordinary setting. The closer pump intensity approaches  the resonance value, the longer becomes the transient period. It may exceed the values of several hundreds times longer than the travel time of the pulse forefront trough the metaslab. A difference between the cases of the  pulsed pump or, alternatively, of the input pulsed signal disappear after the time period about pulse travel time. Reflected pulse begins to develop immediately after pump or seeding pulse edges enter the slab, whereas the pulse in the amplified signal forms with the delay of about the travel time. The outlined coupling scheme and revealed unparalleled properties must be accounted for at  creation of advanced remotely all-optically controlled optical amplifiers, filters, modulators, reflectors and sensors implementing backward electromagnetic waves.

\textbf{Funding.} Ministry of Education and Science of the Russian Federation (No 3.1749.2014/K and 2014/71);
Russian Foundation for Basic Research (RFBR  14-02-00219-a); US Army Research Office (W911NF-14-1-0619).

\end{document}